\documentclass[fleqn,10pt]{wlscirep}
\usepackage[utf8]{inputenc}
\usepackage[T1]{fontenc}
\title{Automatic detection of single-electron regime of quantum dots and definition of virtual gates using U-Net and clustering}

\author[1,2]{Yui Muto}
\affil[1]{Research Institute of Electrical Communication, Tohoku University, 2-1-1 Katahira, Aoba-ku, Sendai 980-8577, Japan}
\affil[2]{Department of Electronic Engineering, Graduate School of Engineering, Tohoku University, Aoba 6-6-05, Aramaki, Aoba-Ku, Sendai 980-8579, Japan}

\author[3, 4]{Michael R. Zielewski}
\affil[3]{Graduate School of Information Sciences, Tohoku University, 6-3-09 Aramaki-aza-Aoba, Aoba-ku, Sendai, 980-8579 Japan}
\affil[4]{Unprecedented-scale Data Analytics Center, Tohoku University, 468-1 Aramaki-aza-Aoba, Aoba-ku, Sendai, 980-8572 Japan}

\author[5]{Motoya Shinozaki}
\affil[5]{WPI Advanced Institute for Materials Research, Tohoku University, 2-1-1 Katahira, Aoba-ku, Sendai 980-8577, Japan}

\author[1,2]{Kosuke Noro}

\author[5, 1, 2, 6, 7,*]{Tomohiro Otsuka}
\affil[6]{Center for Science and Innovation in Spintronics, Tohoku University, 2-1-1 Katahira, Aoba-ku, Sendai 980-8577, Japan}
\affil[7]{Center for Emergent Matter Science, RIKEN, 2-1 Hirosawa, Wako, Saitama 351-0198, Japan}
\affil[*]{tomohiro.otsuka@tohoku.ac.jp}


\begin{abstract}
To realize practical quantum computers, a large number of quantum bits (qubits) will be required. 
Semiconductor spin qubits offer advantages such as high scalability and compatibility with existing semiconductor technologies. 
However, as the number of qubits increases, manual qubit tuning becomes infeasible, motivating automated tuning approaches.
In this study, we use U-Net, a neural network method for object detection, to identify charge transition lines in experimental charge stability diagrams.
The extracted charge transition lines are analyzed using the Hough transform to determine their positions and angles. 
Based on this analysis, we obtain the transformation matrix to virtual gates. Furthermore, we identify the single-electron regime by clustering the Hough transform outputs. 
We also show the single-electron regime within the virtual gate space. These sequential processes are performed automatically. 
This approach will advance automated control technologies for large-scale quantum devices.
\end{abstract}
\begin{document}

\flushbottom
\maketitle

\thispagestyle{empty}

\section*{Introduction}

To advance quantum computing~\cite{2010LaddNat}, the development of quantum bits (qubits) using various systems is progressing. 
Among these, semiconductor spin qubits are one of the promising candidates owing to their high scalability and compatibility with existing semiconductor technologies\cite{koppens2006driven, Maurand2016, Vandersypen2017, Veldhorst2017, Camenzind2022, Zwerver2022, Volk2019}. 
It is estimated that around one million qubits will be required for practical quantum computers\cite{Fowler2012}. 
As the number of qubits increases~\cite{Otsuka2016, Ito2016, Philips2022}, the parameter tuning process becomes complicated and infeasible, making automation critical for large-scale quantum computers\cite{kotzagiannidis2023automated, lennon2019efficiently, Durrer2020}.

To form spin qubits, quantum dots that confine electrons within small regions are used.
It is necessary to control the gate voltages that define the confinement potential of the dots to ensure that each dot contains one electron~\cite{1996TaruchaPRL, koppens2006driven}. 
One of the challenges in this process is an unintended interaction between quantum dots and gate electrodes. 
For example, while each gate voltage is designed to control the electron number within the quantum dot corresponding to its electrode, the electron number in neighboring quantum dots may also be affected.
Such an unintended interaction complicates tuning, as the interdependency between gate voltages must be considered.
One tool that enables independent control of the electron number within each quantum dot is virtual gates\cite{Volk2019, Hsiao2020}. 
These gates can be constructed by measuring the effect of the gate voltages on neighboring dots.  
This can be initially achieved by slightly varying the voltage of each gate electrode and analyzing the shift of the charge transition point\cite{Volk2019}. 
However, this process is infeasible to perform manually for every gate electrode, especially in future large-scale qubits.

To address this issue, automated approaches have been developed to define virtual gates using image processing techniques\cite{Van2018, Mills2019, Oakes2020, Liu2022, Ziegler2023, ziegler2023tuning}. 
A key technique is the Hough transform, which automatically detects charge transition lines (CT-lines) in charge stability diagrams (CSDs)\cite{Lapointe2020}. 
While the Hough transform requires binarized images, extracting CT-lines from images processed using common binarization methods, such as the Canny edge detection and Otsu’s methods, remains challenging. This difficulty arises through noise introduced by the device and experimental conditions~\cite{ziegler2022toward, Hader2024Sim}, which persists through binarization, and leads to unexpected results when applying the Hough transform. 
Therefore, to obtain high-precision virtual gates using the Hough transform, a method is needed that is robust to noise and can accurately extract only the CT-lines within the CSDs.

With recent advancements in the field of machine learning, automated object detection methods have been developed and applied across various domains. One such method is U-Net\cite{ronneberger2015u}, a segmentation model originally developed for biomedical applications\cite{Siddique2021U-Net}. By training U-Net on annotated data of the target objects within images, precise automated segmentation can be achieved\cite{Hader2024}.

In this study, we develop a method for the automatic definition of virtual gates and detection of the single-electron regime (SER).
We employ a U-Net model to automatically detect CT-lines in CSD and apply the Hough transform to analyze these detected lines. Using this information, we define virtual gates. 
Additionally, we compare the accuracy of the binarization and Hough transform methods with alternative methods. 
We also identify the transition line positions using clustering and find the leftmost bottom crossing of the CT-lines adjacent to the SER, which is an important state for semiconductor spin qubits. 

\section*{Results and discussion}
\subsection*{Accuracy verification of trained U-Net model}

\begin{figure*}[ht]
\centering
\includegraphics[width=\textwidth]{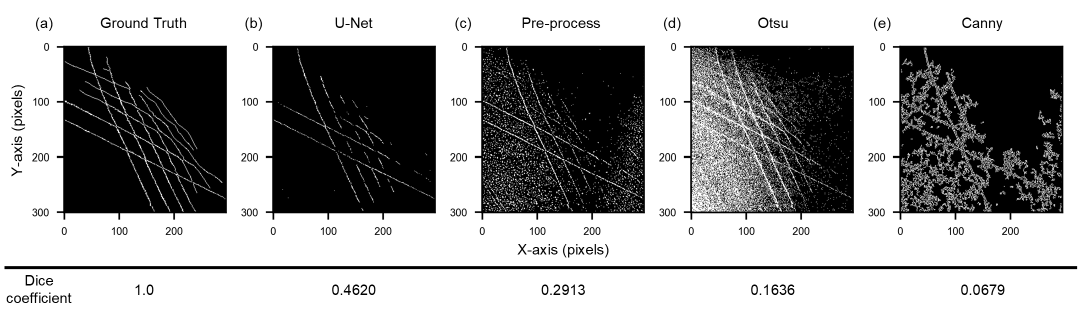}
\caption{Binarized CSD images of (a) Ground truth and those processed by (b) U-Net, (c) Pre-process, (d) Otsu's method, and (e) Canny edge detection.
The values indicate the Dice coefficient of each method. The value closer to 1 indicates a higher similarity between the detection result and the ground truth.}
\label{fig1}
\end{figure*}

We apply the trained U-Net model to CSD of double quantum dots. 
The details of the training are described in the Methods section. 
The U-Net outputs pixel-wise segmentation probabilities for background and CT-lines, which are binarized by assigning the class with higher probability (1 for CT-lines, 0 for background).
To compare with conventional methods, three other methods (pre-process, Otsu's method, and Canny edge detection) are also tested. 
The pre-process method, proposed in our previous paper\cite{muto2024}, combines techniques such as Gaussian filtering to suppress noise in the CSD before binarizing the image.
Otsu's method\cite{Otsu1979, tensmeyer2020historical} is a well-known technique that automatically determines the threshold value for binarizing the image. 
While the Canny method\cite{Canny1986} typically requires manual threshold setting, it is commonly used as a binarization technique for edge detection before applying the Hough transform.
We utilize the OpenCV package to apply both Otsu's method\cite{opencv_otsu_binarization} and the Canny method\cite{opencv_canny_method}.

Figure~\ref{fig1}(a) shows the manually annotated ground truth data, and Figs.~\ref{fig1}(b)–(e) display the results of binarization in the CSD using each method. 
The values below each figure represent the Dice coefficient, where a score closer to 1 indicates better alignment with the ground truth segmentation. 
As seen from these results, U-Net achieves the better alignment and extracts only CT-lines as binarized pixels.
In contrast, the other three methods (Figs.~\ref{fig1}(c)–(e)) pick up experimental noise, reflecting low scores. 
This performance reflects the ability of U-Net to perform segmentation-based probabilistic classification, in which its skip connections allow simultaneous evaluation of both pixel-level features and global line continuity.
By training on experimental data, we achieve robust classification even under noisy experimental conditions.
Note that some CT-lines in the upper-right region of the CSD are not detected by U-Net. 
In this region, the signal is weak, making it difficult to distinguish CT-lines even for human experts. 
This could potentially be addressed by device and measurement developments such as techniques to maintain optimal charge sensor sensitivity~\cite{Hader2023noise, fujiwara2023}.

\subsection*{Hough transform and definition of virtual gate}
\begin{figure*}[ht]
\centering
\includegraphics[width=\textwidth]{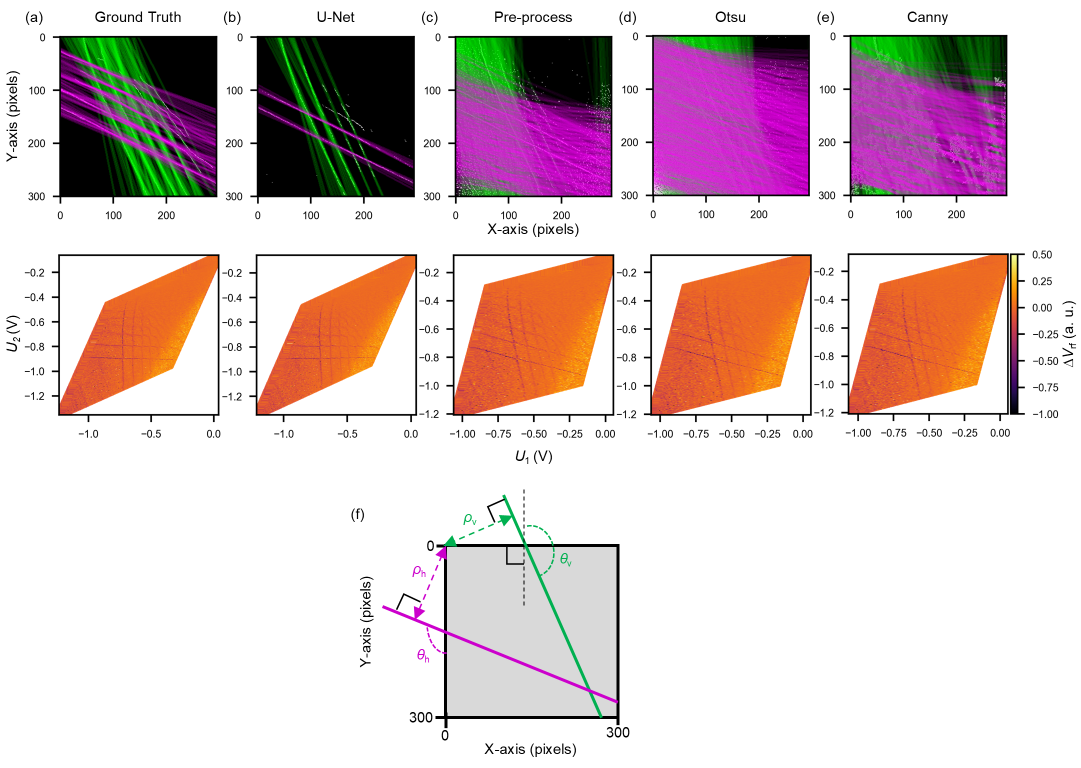}
\caption{Detection of angle and position of CT-lines using the Hough transform for (a) ground truth, (b) U-Net, (c) pre-process, (d) Otsu's method, and (e) Canny edge detection. 
Green and pink lines indicate vertical-like and horizontal-like CT-lines, respectively. 
Lower row shows charge stability diagrams with automatically defined virtual gate axes.
(f) Schematic of the output parameters from the Hough transform}
\label{fig2}
\end{figure*}

The Hough transform is applied to the binarized CSD images obtained from each method to detect the angle and position of the CT-lines. 
These output parameters are used to automatically define the transformation matrix for virtual gates.

In a double quantum dot system, two types of CT-lines appear in the CSD: those that are nearly vertical and horizontal. 
In the case of multiple dot systems, additional CT-lines at different angles will appear.
Virtual gates can reduce these CT-lines to two types by enabling independent control of the charge state in each dot.

The Hough transform is implemented using the OpenCV package\cite{opencv_hough_lines}. 
The key parameters used in this study are summarized in Table~\ref{table1}.
\begin{table}[h!]
    \centering
    \caption{Parameters used in the \texttt{cv2.HoughLines} function}
    \begin{tabular}{|c|c|c|}
        \hline
        \textbf{Parameter} & \textbf{vertical} & \textbf{horizontal}\\
        \hline
        \texttt{rho} & 1 & 1 \\
        \hline
        \texttt{theta} & $\pi/180$ & $\pi/180$ \\
        \hline
        \texttt{threshold} & 30 & 30 \\
        \hline
        \texttt{min\_theta} & $5\pi/6$ & $\pi/2$ \\
        \hline
        \texttt{max\_theta} & $\pi$ & $2\pi/3$ \\
        \hline
    \end{tabular}
    \label{table1}
\end{table}
Here, \texttt{rho} is the distance resolution of the accumulator in pixels, \texttt{theta} is the angle resolution of the accumulator in radians, \texttt{threshold} is the minimum number of votes required for a line to be considered as a valid straight line, \texttt{min\_theta} is the minimum angle to check for lines, and \texttt{max\_theta} is the upper bound for the angle. 
The Hough transform outputs the parameters $\rho$ and $\theta$ of the detected lines, where $\rho$ represents the perpendicular distance from origin to the line, and $\theta$ represents the angle formed by this perpendicular line and the horizontal axis measured in counter-clockwise.

We apply the Hough transform to the binarized CSD images from each method and show the results in the upper row of Figs.~\ref{fig2}(a)–(e). 
These figures are prepared by overlaying the straight lines detected by the Hough transform onto the binarized images shown in Fig.~\ref{fig1}. 
The green lines represent the vertical-like CT-lines, and the pink lines represent the horizontal-like CT-lines. 
The same parameters for the Hough transform are used for all methods. 

As seen in the upper row of Fig.~\ref{fig2}(a), accurately segmented CT-lines provide reliable angle and position values through the Hough transform.

Almost all of the CT-lines detected by U-Net are also captured by the Hough transform as shown in Fig.~\ref{fig2}(b). 
Note that some of the horizontal-like CT-lines are not detected. 
This is because unclear CT-lines constructed by a few pixels, such as those caused by sensor sensitivity issues, do not meet the threshold required by the Hough transform.

On the other hand, the Hough transform detects many lines other than CT-lines in the images binarized by other methods, as shown in the upper row of Figs.~\ref{fig2}(c)–(e)).
One of the possible scenarios is that sequential noise pixels are detected as CT-lines, as observed in Figs.~\ref{fig1}(c)-(e)).
Such misclassification due to noise pixels was also reported in a previous study\cite{muto2024}.

To automatically define the virtual gate $\mathbf{U}$, we use the angle $\theta$ from the Hough transform. 
We define the virtual gate $\mathbf{U}$ using the mean angle $\theta$ calculated for both vertical-like and horizontal-like CT-lines, as follows:
\begin{equation}
\mathbf{U} = \mathbf{G} \cdot \mathbf{V}, 
\end{equation}
\begin{equation}
    \mathbf{U} = \begin{bmatrix}
        U_{1} \\
        U_{2}
        \end{bmatrix}, 
\end{equation}
\begin{equation}
    \mathbf{G} = \begin{bmatrix}
        -\cos \theta_{\text{v}} & \sin \theta_{\text{v}} \\
        -\cos \theta_{\text{h}} & \sin \theta_{\text{h}}
        \end{bmatrix}.
\end{equation}
\begin{equation}
    \mathbf{V} = \begin{bmatrix}
        V_{g1} \\
        V_{g2}
        \end{bmatrix}, 
\end{equation}
Here, $\mathbf{V}$ is the physical plunger gate, $\theta_{\text{v}}$ is the mean angle $\theta$ of the lines detected as vertical-like CT-lines, and $\theta_{\text{h}}$ is the mean angle $\theta$ of the lines detected as horizontal-like CT-lines, as illustrated in Fig.~\ref{fig2}(f). 

The CSDs transformed by virtual gates are shown in the lower row of Fig.~\ref{fig2}. 
In the case of the ground truth data (Fig.~\ref{fig2}(a)), it works well as the CT-lines appear at right angles. 
The virtual gates utilizing U-Net (Fig.~\ref{fig2}(b)) display similar transformed CSDs to the ground truth because U-Net accurately extracts only the CT-lines from the CSD.
For the other methods (Figs.~\ref{fig2}(c)-(e)), the virtual gates do not work well.
It is difficult to avoid the effect of additional noise lines on errors in $\theta_{\text{v}}$ and $\theta_{\text{h}}$, resulting in poorly defined virtual gates.
These results show that U-Net can accurately detect CT-lines from noisy CSD data and is useful for automatically defining virtual gates through the Hough transform.

\subsection*{Identification of CT-lines by clustering}
\begin{figure}[ht]
\centering
\includegraphics[width=\textwidth]{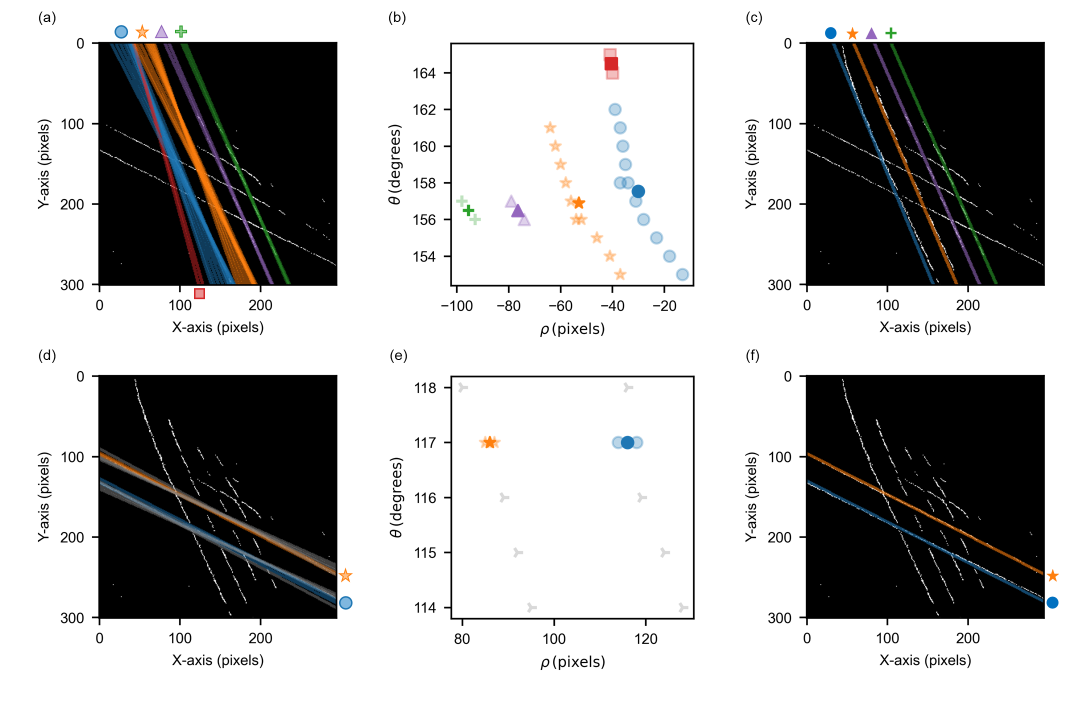}
\caption{(a) Identification of vertical-like CT-lines by clustering. (b) Clustering results in the $\theta$-$\rho$ space, where each color corresponds to a cluster of detected CT-lines. (c) Merged lines obtained from multiple lines within each cluster. (d)-(f) Same analysis for horizontal-like CT-lines. 
Gray lines and markers represent noise pixels identified by DBSCAN, which are not assigned to any cluster.}
\label{fig3}
\end{figure}

As shown in Fig.~\ref{fig2}(b), the Hough transform detects multiple lines for a single CT-line, making it difficult to find SER although virtual gates can be defined.
To address this issue, we apply clustering to the detected lines to identify each individual CT-line.

For the clustering process, we apply Density-based Spatial Clustering of Applications with Noise (DBSCAN)\cite{sklearn_DBSCAN} to the standardized $\rho$ and $\theta$. 
The standardization process converts the data to have a zero mean and a standard deviation of 1.
DBSCAN identifies clusters by grouping data points that are close to each other, enabling cluster detection even in complex scatter patterns.
This method does not require specifying the number of clusters in advance, which is advantageous for automation.
The clustering results of $\rho$ and $\theta$ obtained from the Hough transform of vertical-like CT-lines are shown in Figs.~\ref{fig3}(a) and (b).
The clustering successfully classifies most of the detected lines.

We identify each individual CT-line by taking the average of $\rho$ and $\theta$ in each cluster. The thick markers in Fig.~\ref{fig3}(b) show these averages.
Sometimes, the number of obtained clusters is greater than the actual CT-lines, resulting in an extra cluster and an averaged line (shown as red square markers in Figs.~\ref{fig3}(a) and (b)).
To remove this extra line, we use the following approach.
Since CT-lines are nearly parallel to each other, they rarely intersect.
Thus, if such an intersection is detected, it indicates duplicate detection of the same CT-line.
In this case, we eliminate the extra line (the red marker) whose $\theta$ value deviates most from the mean $\theta$ of all lines.

As seen in Fig.~\ref{fig3}(c), the proposed clustering and merging process extracts all the CT-lines segmented by U-Net.
This process also works well for horizontal-like CT-lines, as demonstrated in Figs.~\ref{fig3}(d)-(f).

\subsection*{Finding of the single electron regime in the charge stability diagram}
\begin{figure}[ht]
\centering
\includegraphics[width=\textwidth]{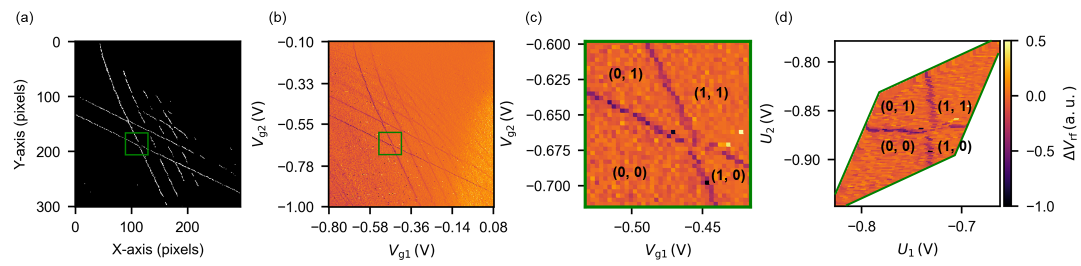}
\caption{Automatically detected single-electron regime (green rectangle) in charge stability diagrams: (a) binarized map and (b) experimental data. (c) Zoomed-in view of single-electron region with gate voltage axes and (d) with virtual gate axes.}
\label{fig4}
\end{figure}

Finally, we identify the SER in the CSD using the extracted CT-lines.
We calculate the intersection point between the leftmost line (blue) in Fig.~\ref{fig3}(c) and the bottommost line (blue) in Fig.~\ref{fig3}(f).
This intersection corresponds to the leftmost bottom crossing of the CT-lines, marked by the automatically determined green rectangle in Fig.~\ref{fig4}(a).
Fig.~\ref{fig4}(b) shows the experimental data with the marked SER region, and Fig.~\ref{fig4}(c) provides a zoomed-in view of this region highlighted by the green rectangle.
Here, ($n$, $m$) indicates the number of electrons in each quantum dot.
Furthermore, we display this SER using virtual gates, as shown in Fig.~\ref{fig4}(d).
These results demonstrate fully automated detection of CT-lines in CSD and finding SER with virtual gates.

\begin{figure}[ht]
\centering
\includegraphics[width=\textwidth]{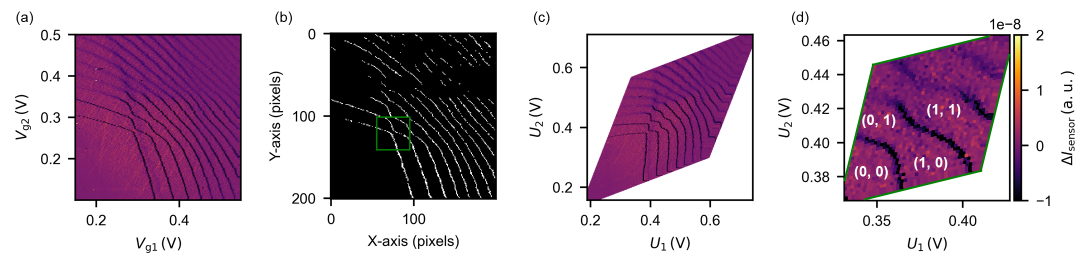}
\caption{(a) Charge stability diagram from another group\cite{zwolak2022qflow} with differential processing along the horizontal axis. (b) Output of trained U-Net model with identified single-electron regime (green rectangle). (c) Charge stability diagram visualized with automatically defined virtual gates. (d) Single-electron regime represented in virtual gate space.}
\label{fig5}
\end{figure}

To validate the versatility of our approach, we apply our method to data from another group~\cite{zwolak2022qflow}.
Fig.~\ref{fig5} shows the raw data with differential processing along the horizontal axis and results of the processing flow we proposed.
We also demonstrate the successful binarization of CSD using U-Net, definition of virtual gates, and finding of SER in CSD for another group's data, showing the robust performance of our method.

\section*{Conclusion}
In this study, we demonstrate a method for the automatic definition of virtual gates and detection of the SER.
We use U-Net to extract CT-lines from noisy CSD data and determine their angle and position by the Hough transform.
This enables us to automatically define virtual gates.
We also identify transition line positions using DBSCAN clustering and merging processes.
Based on this approach, we demonstrate automatic display of SER in CSD with virtual gate axes.
Furthermore, we apply our method to data from another group and confirm that it works well.
Our results are expected to contribute to automatic quantum dot analysis and tuning, advancing the development of large-scale quantum devices with multiple quantum dots.

\section*{Methods}
\subsection*{Training of U-Net model}

\begin{figure}[ht]
\centering
\includegraphics[width=\textwidth]{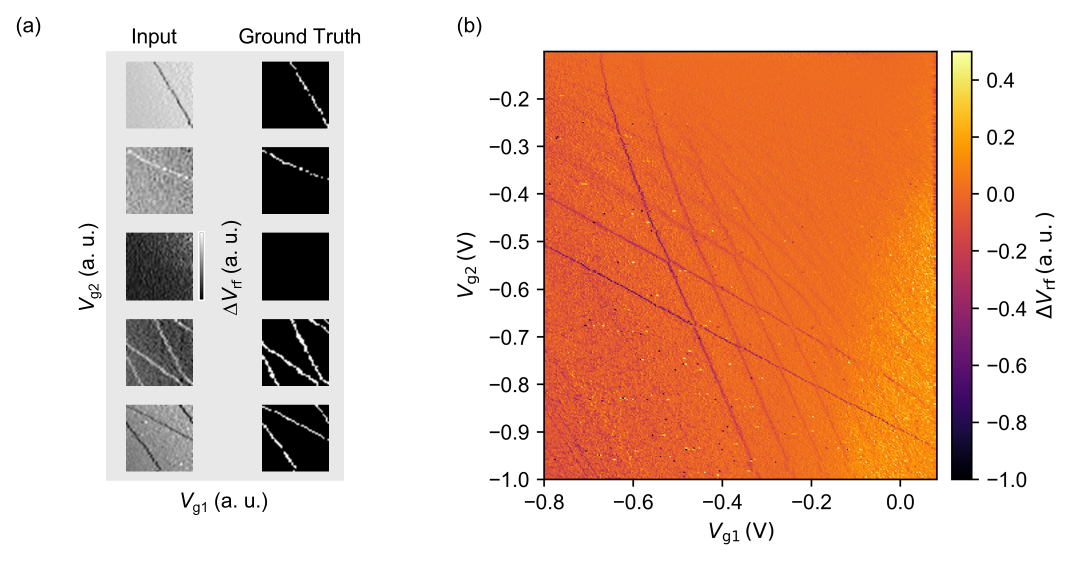}
\caption{(a) Examples of training data pairs for the U-Net model, showing input data and corresponding ground truth (43 × 43 pixels). (b) A 301 × 295 pixel charge stability diagram used for evaluating the trained U-Net model, with differential processing along the horizontal axis.}
\label{fig6}
\end{figure}

We describe the dataset used for training the U-Net model.
Initially, we prepare 11 CSD images measured over a wide range of voltages from several different quantum devices.
Each CSD is numerically differentiated along the horizontal axis to enhance the visibility of CT-lines, which are then used as input data for the U-Net model.
The ground truth data is prepared manually by tracing the CT-lines.
During training, random regions of size 48 × 48 pixels are extracted from the 11 images and downscaled to 43 × 43 pixels for the U-Net model input.
Figure~\ref{fig6}(a) shows examples of the training data, where the left and right rows represent the input and ground truth data, respectively.
A total of 182,101 images are used for training, of which 10\% are used for validation.

To enhance the accuracy of CT-line detection, we employ two supporting techniques: "Random Invert"\cite{torchvision_randominvert}, which randomly inverts the colors of the given image, and "Random Adjust Gamma", which randomly adjusts the contrast of the image using gamma correction\cite{torchvision_adjust_gamma}.
Random Invert is incorporated to enable the U-Net model to handle various CT-line colors. 
Since the measured CT-lines may appear either brighter or darker relative to the background depending on the sensor quantum dot condition, this technique improves the model's adaptability to various CT-line appearances.
Similarly, Random Adjust Gamma ensures that the U-Net model can handle CSDs with varying contrast levels, which change with sensor sensitivity. This technique enhances the model's robustness in detecting CT-lines under various sensor conditions.
These techniques are randomly applied to training images in arbitrary combinations, or not at all.

In this study, we partially modify a pre-existing U-Net library\cite{milesial_pytorchunet} and use it for our model training. 
We adopt the original U-Net architecture and implement the model in PyTorch (v.2.0.1).
The hyperparameters are set as follows: a batch size of 64, a learning rate of 0.00001, and 100 epochs.
Note that the actual number of epochs during training is 66 due to the early stopping to prevent overfitting.

The performance of the U-Net model is evaluated by the Dice coefficient. 
For a predicted result \(\mathbf{X}\) and ground truth data \(\mathbf{Y}\), the Dice coefficient \( \mathrm{Dice}(\mathbf{X}, \mathbf{Y}) \) is defined as follows~\cite{dice1945measures}:
\begin{equation}
\mathrm{Dice}(\mathbf{X}, \mathbf{Y}) = \frac{2 \sum_{i} X_i Y_i}{\sum_{i} X_i + \sum_{i} Y_i}.
\end{equation}
Here, \(X_i\) and \(Y_i\) represent the predicted and ground truth pixel values respectively, both taking binary values of 0 or 1.
The Dice coefficient is a metric used to evaluate the overlap between the predicted result and the ground truth data, taking values between 0 and 1. 
A value closer to 1 indicates that the binarization result is similar to the ground truth.

The Dice loss, a loss function that should be minimized, is defined as follows:
\begin{equation}
\mathrm{Dice Loss} = 1 - \mathrm{Dice}(\mathbf{X}, \mathbf{Y}).
\end{equation}
The Dice loss is used to train the model to maximize the Dice coefficient, which increases the overlap between the predicted result and the ground truth.
After each training epoch, we calculate the average Dice loss of the validation data over all batches.
If there is no improvement for 5 consecutive epochs, training is stopped, and the model with the lowest Dice loss is selected.

The performance of the trained U-Net model is evaluated using the experimental data~\cite{riec_data} shown in Fig.~\ref{fig6}(b). 
This figure is processed with horizontal differentiation and has a size of 301 × 295 pixels.

\bibliography{references}

\section*{Acknowledgements}
We thank 
RIEC Fundamental Technology Center and the Laboratory for Nanoelectronics and Spintronics
for the technical support. 
Part of this work is supported by 
MEXT Leading Initiative for Excellent Young Researchers, 
Grants-in-Aid for Scientific Research (23KJ0200, 21K18592, 23H01789, 23H04490), 
FRiD Tohoku University.
Y. M. acknowledges WISE Program of AIE for financial support.

\section*{Author contributions}
Y. M. and T. O. planned the project; 
Y. M. and M. R. Z. developed the U-Net model; 
Y. M. developed the clustering method;
Y. M., M. S., K. N., and T. O. conducted data analysis; 
all authors discussed the results;
Y. M., M. S., and T. O. wrote the manuscript with inputs from all authors.

\end{document}